\documentclass{emulateapj}
\usepackage{graphicx,natbib,amsmath,slashbox,soul} 
\usepackage{CJKutf8}

\newcommand{\Lsun}{$L_{\odot}$}
\newcommand{\LIR}{$L_{\rm IR}$}  
\newcommand{\mum}{$\mu m$}
\newcommand{\Mstar}{$M_{*}$}

\newcommand{\Mtwenty}{$M_{20}$}

\newcommand{\uft}{$u^{\prime}$}
\newcommand{\gft}{$g^{\prime}$}
\newcommand{\rft}{$r^{\prime}$}
\newcommand{\ift}{$i^{\prime}$}
\newcommand{\zft}{$z^{\prime}$}

\newcommand{\lameff}{${\rm \lambda_{eff}}$}
\newcommand{\dlam}{${\rm \Delta \lambda}$}

\begin{document}

\title{A comparison of the morphological properties between local and $z\sim1$ infrared luminous galaxies. Are local and high$-z$ (U)LIRGs different?}

\author{Chao-Ling Hung \begin{CJK*}{UTF8}{bsmi}(洪肇伶)\end{CJK*}\altaffilmark{1,2,3}}
\author{D. B. Sanders\altaffilmark{1}}
\author{Caitlin M. Casey\altaffilmark{4}}
\author{Michael Koss\altaffilmark{5}}
\author{Kirsten L. Larson\altaffilmark{1}}
\author{Nicholas Lee\altaffilmark{1}}
\author{Yanxia Li\altaffilmark{1}}
\author{Kelly Lockhart\altaffilmark{1}}
\author{Hsin-Yi Shih\altaffilmark{1}}
\author{Joshua E. Barnes\altaffilmark{1}}
\author{Jeyhan S. Kartaltepe\altaffilmark{6}}
\author{Howard A. Smith\altaffilmark{3}}

\affil{\altaffilmark{1} Institute for Astronomy, University of Hawaii, 2680 Woodlawn Drive, Honolulu, HI 96822, USA; clhung@ifa.hawaii.edu}
\affil{\altaffilmark{2} SAO Pre-doctoral Fellow}
\affil{\altaffilmark{3} Harvard-Smithsonian Center for Astrophysics, Cambridge, MA 02138, USA}
\affil{\altaffilmark{4} Department of Physics and Astronomy, University of California at Irvine, 2162 Frederick Reines Hall, Irvine, CA 92697, USA}
\affil{\altaffilmark{5} Institute for Astronomy, ETH Z\"{u}rich, Wolfgang-Pauli-Strasse 27, CH-8093 Z\"{u}rich, Switzerland}
\affil{\altaffilmark{6} National Optical Astronomy Observatory, 950 North Cherry Ave., Tucson, AZ, 85719, USA}

\begin{abstract}
Ultraluminous and luminous infrared galaxies (ULIRGs and LIRGs) are the most extreme star-forming galaxies in the universe, and dominate the total star formation rate density at $z>1$.
In the local universe ($z<0.3$), the majority of ULIRGs and a significant portion of LIRGs are triggered by interactions between gas-rich spiral galaxies, yet it is unclear if this is still the case at high$-z$. 
To investigate the relative importance of galaxy interactions in infrared luminous galaxies, we carry out a comparison of optical morphological properties between local (U)LIRGs and (U)LIRGs at $z=0.5-1.5$ based on the same sample selection, morphology classification scheme, and optical morphology at similar rest-frame wavelengths.
In addition, we quantify the systematics in comparing local and high$-z$ datasets by constructing a redshifted dataset from local (U)LIRGs, in which its data quality mimics the high$-z$ dataset.
Based on the {\it Gini}-\Mtwenty\ classification scheme, we find that the fraction of interacting systems decreases by $\sim$ 8\% from local to $z\lesssim1$, and it is consistent with the reduction between local and redshifted datasets ($6^{+14}_{-6}\%$).
Based on visual classifications, the merger fraction of local ULIRGs is found to be $\sim20\%$ lower compared to published results, and the reduction due to redshifiting is $15^{+10}_{-8}\%$.
Consequently, the differences of merger fractions between local and $z\lesssim1$ (U)LIRGs is only $\sim$ 17\%.
These results demonstrate that there is no strong evolution in the fraction of (U)LIRGs classified as mergers at least out to $z\sim1$.
At $z>1$, the morphology types of $\sim$ 30\% of (U)LIRGs can not be determined due to their faintness in the $F814W$-band, and thus the merger fraction measured at $z>1$ suffers from large uncertainties.    

\end{abstract}


\section{Introduction}

Ultraluminous infrared galaxies (ULIRGs; \LIR\footnote{\LIR\ $\equiv L_{8-1000\mu m}$ in the object's rest-frame}\ $\geq10^{12}$\Lsun) and Luminous Infrared Galaxies (LIRGs; $10^{11}\leq$ \LIR\ $<10^{12}$\Lsun) host the most intense star formation events in the universe. 
Locally ($z\lesssim0.3$), the majority of ULIRGs and a significant portion of LIRGs are triggered by strong interactions between gas-rich spiral galaxies, where the bulk of the infrared emission is produced from dust heating by starbursts and/or an active galactic nucleus (AGN) \citep[e.g.][]{Sanders1988,Sanders1996}.
Detailed morphological and kinematic studies of (U)LIRGs have revealed various interacting features such as bridges, tidal tails, and complicated kinematic properties \citep[e.g.][]{Mihos1998,Farrah2001,Veilleux2002,Colina2005}.
The strong correlation between galaxy interactions and (U)LIRGs can also be seen via the growing merger fraction and the progressing merger stage with increasing \LIR\ \citep[e.g.][Larson et al. in prep.]{Veilleux2002,Ishida2004,Haan2011,Ellison2013}.
Extensive simulation work has also demonstrated that galaxy interactions can transport gas toward the merger nuclei where it can trigger enhanced star formation and AGN activities \citep[e.g.][]{Barnes1996,Mihos1996}.

Although (U)LIRGs are relatively rare locally, they become increasingly important contributors to the total star formation rate density at high$-z$ \citep[e.g. $\gtrsim70\%$ at $z\sim1$,][]{Le-Floch2005,Casey2012a}.
However, it remains unclear if galaxy interactions are the dominant mechanism driving intense star formation in high$-z$ (U)LIRGs.
In fact, high$-z$ star-forming galaxies are known to contain a higher fraction of gas compared to their local counterparts \citep{Tacconi2008,Tacconi2010}. 
Such enhanced gas fractions may increase the luminosity threshold between normal star formation and merger-driven starbursts \citep{Hopkins2010}.
In these scenarios, high$-z$ (U)LIRGs may be alternatively driven by secular accretion as opposed to galaxy interactions.

To investigate the role of galaxy interactions in high$-z$ (U)LIRGs and to understand how the threshold between normal and merger-driven star formation varies across cosmic time, one can explore the frequency of mergers in these extreme star-forming systems and compare such frequencies at multiple epochs.
Extensive studies have already been done via the characterization of rest-frame optical morphology \citep[e.g.][]{Lotz2008a,Dasyra2008,Melbourne2008,Kartaltepe2010a,Kartaltepe2012,Hung2013}.
Although the percentage of mergers in $z\sim1$ LIRGs can vary from $\sim40$\% \citep{Kartaltepe2010a,Hung2013} to only 10$-$15\% \citep{Lotz2008a,Melbourne2008}, there is a general trend that the merger fraction in high$-z$ (U)LIRGs appears to be significantly lower compared to their local counterparts.
For example, the merger fraction in ULIRGs found at $z\sim1$ and $z\sim2$ is $\sim50\%$ \citep{Kartaltepe2010a,Kartaltepe2012,Hung2013}, whereas a much higher merger fraction is found in local ULIRGs \citep[$>85\%$,][]{Farrah2001,Veilleux2002,Ishida2004}.

The comparison between local and high$-z$ datasets is often not straightforward.
Two major factors influencing such comparisons are surface brightness dimming and band-shifting.
Deep optical surveys of high$-z$ galaxies correspond to rest-frame $B$-band or shorter, yet galaxies may show very different structures when imaged at longer wavelengths \citep[e.g.][]{Cameron2011}.
Locally, high resolution/sensitivity imaging at rest-frame $I$-band or longer wavelength can detect large-scale tidal features and other disturbed structures, enabling an unambiguous identification of merger signatures \citep[e.g.][]{Surace1998,Veilleux2002}.
Yet these extended features often fade away or become less prominent at shorter wavelengths or when artificially redshifted to high$-z$ \citep{Hibbard1997,Petty2009}.
A proper consideration of the reduction of merger fractions that is purely due to data degradation and band-shifting is important when comparing local and high$-z$ datasets.

Another complication may originate from the sample identification of (U)LIRGs. 
The majority of local (U)LIRGs have been identified from the {\it Infrared Astronomical Satellite (IRAS)} 60 \mum\ All-Sky Survey \citep[e.g.][]{Kim1998,Lawrence1999,Sanders2003}.
However, prior to the availability of observations from the {\it Herschel Space Observatory} \citep{Pilbratt2010}, the selection of high$-z$ (U)LIRGs often relied on extrapolation from {\it Spitzer} 24 \mum\ observations (which corresponds to rest-frame 8 \mum\ at $z\sim2$) or optical/near-infrared color selection \citep{Le-Floch2004,Daddi2004}.
Selections based on mid-IR emission can bias toward galaxies with a stronger AGN \citep[e.g.][]{Brand2006}, and it is unclear how this may impact the statistics of galaxy morphology such as merger fractions.

To avoid sample selection biases, we compile samples of (U)LIRGs at $z\sim0$ and $z=0.5-1.5$ that are both selected based on rest-frame 60$-$100 \mum\ observations.
We carry out a comparison of their morphology using both automatic indicators ($Gini$ coefficient and \Mtwenty) and visual classifications.
Since our high$-z$ sample is observed at rest-frame $U$- and $B$-band, we limit our comparison to those local (U)LIRGs with observations at similar wavelengths.
As a test, we take nearby (U)LIRGs and simulate how they would appear at high$-z$.
This way we can directly test how the surface brightness dimming influences any signature of merging.

The paper is organized as follows: the data and sample selection are described in Section 2 and we construct the redshifted dataset in Section 3.
We describe our classification schemes in Section 4 and then examine the distribution of galaxy morphological properties in three \LIR\ bins in Section 5.
We discuss the implications of our results in Section 6 and summarize our findings in Section 7.
Throughout this paper, we adopt a $\Lambda$CDM cosmology with $H_0=70$ km s$^{-1}$ Mpc$^{-1}$, $\Omega_{M}=0.3$ and $\Omega_{\Lambda}=0.7$ \citep{Hinshaw2009}.
Magnitudes are given in the AB system.

\section{Sample and Data for IR-luminous Galaxies}

\subsection{Sample Selection}

Our sample of local (U)LIRGs has been selected from the Great Observatories Allsky LIRGs Survey \citep[GOALS,][]{Armus2009}, and the QDOT all-sky {\it IRAS} galaxy redshift survey \citep[Queen Mary and Westfield College, Durham, Oxford and Toronto;][]{Lawrence1999}.
Both GOALS and QDOT samples are selected based on {\it IRAS} 60 \mum\ observations.
The GOALS sample comprises 203 (U)LIRGs with \LIR$>10^{11}$ \Lsun\ at $z<0.088$, and it is a complete subset of the {\it IRAS} Revised Bright Galaxy Sample \citep[RBGS,][]{Sanders2003}, which includes all extragalactic sources with 60 \mum\ flux density $>$ 5.24 Jy.
The QDOT sample is selected based on a 60 \mum\ flux density limit of 0.6 Jy, which provides 95 (U)LIRGs with \LIR$>10^{11.94}$ \Lsun\footnote{Note that the luminosity listed in \citet{Lawrence1999} is determined using 60 \mum\ flux. \LIR\ of these QDOT ULIRGs are determined based on the method described in \citet{Kim1998}. \LIR$=4\pi D_L^2\times F_{\rm IR}$, where the infrared flux $F_{\rm IR}$[$10^{-14}$ W m$^{-2}$]$=1.8\times(13.48 \times f_{12\mu m} + 5.16 \times f_{25\mu m} +  2.58 \times f_{60\mu m} +  1.00 \times f_{100\mu m})$.} at $z<0.34$.
 
The sample of (U)LIRGs at $z=0.5-1.5$ is selected based on {\it Herschel} Photodetector Array Camera and Spectrometer \citep[PACS;][]{Poglitsch2010} and the Spectral and Photometric Imaging REceiver \citep[SPIRE;][]{Griffin2010} observations in the Cosmic Evolution Survey \citep[COSMOS;][]{Scoville2007}.
The PACS (100 and 160 \mum) and SPIRE (250, 350 and 500 \mum) observations in the COSMOS field are acquired as part of the PACS Evolutionary Probe program \citep[PEP;][]{Lutz2011} and the Herschel Multi-tiered Extragalactic Survey \citep[HerMES;][]{Oliver2012}, respectively.
Detailed descriptions of the source identification, the cross-correlation between multi-wavelength datasets, and the measurement of \LIR, dust temperature and dust mass are presented in \citet{LeeN2013}, where they identify a sample of 4218 {\it Herschel}-selected sources with $z=0.02-3.54$ and log(\LIR/\Lsun) $=9.4-13.6$.

These {\it Herschel}-selected galaxies, similar to the local {\it IRAS}-selected (U)LIRGs, are identified at rest-frame $\sim60-100$\mum.
Such Far-infrared (FIR) selection identifies IR-luminous galaxies at the peak of their SEDs, which is less biased by the emission from AGNs and does not rely on the extrapolation from mid-IR data.
Based on the {\it Herschel}-selected galaxies identified by \citet{LeeN2013}, \citet{Hung2013} have carried out a detailed morphological analysis of 2084 galaxies at $0.2<z<1.5$ using the Advanced Camera for Surveys (ACS) $F814W$-band (\lameff\ = 7985\AA, \dlam\ = 1877\AA) images taken as part of the {\it HST}-COSMOS survey \citep{Koekemoer2007}.
In this paper, we select a subset of 246 (U)LIRGs at $z=0.5-1.5$ with log(\LIR/\Lsun) $=11.0-12.7$ (hereafter the high$-z$ sample, note that this term specifically refers to our {\it Herschel}-selected (U)LIRGs at $0.5<z<1.5$), where this subset is uniformly distributed in \LIR\ but otherwise randomly selected.
 
\subsection{Optical Images}

The processed ACS $F814W$-band images used for our high$-z$ sample have the pixel scale of 0.03 \arcsec\ pixel$^{-1}$ and the point spread function (PSF) of 0.1\arcsec\ \citep{Koekemoer2007}, which correspond to the physical scale of $\sim250$ pc pixel$^{-1}$ and $\sim800$ pc at $z=1$.
For the local sample, we use a subset of GOALS galaxies that are covered in the Sloan Digital Sky Survey Data Release 7 \citep[SDSS DR7:][hereafter the GOALS-SDSS sample]{Abazajian2009}. 
SDSS images have a complete wavelength coverage in \uft, \gft, \rft, \ift\ and \zft\ filters (\lameff\ = 3595, 4640, 6122, 7439, 8897\AA, \dlam\ = 558,  1158, 1111, 1044, 1124\AA), and thus we can properly consider K-correction when constructing the redshifted datasets.
The pixel scale and the typical PSF of SDSS images are 0.396\arcsec\ pixel$^{-1}$ and 1.3\arcsec\ \citep{Abazajian2009}, which corresponds to the physical scale of $\sim240$ pc pixel$^{-1}$ and  $\sim800$ pc at $z=0.03$, the median redshift of the GOALS-SDSS galaxies.
We have restricted the GOALS-SDSS sample to the galaxies that are not saturated in the SDSS images, and those are at $z\lesssim0.06$ which have sufficient spatial resolution for simulating high$-z$ {\it HST} observations.
We have also excluded two fields that contain too many stars because of the difficulties in removing the field stars from target galaxies.
The final GOALS-SDSS sample contains 59 galaxies with log(\LIR/\Lsun) $=11.0-12.4$.

In addition to the GOALS-SDSS sample, we also include 23 (U)LIRGs with log(\LIR/\Lsun) $=11.9-12.9$ from the QDOT survey that have been observed with the {\it HST} Wide-Field Planetary Camera 2 (WFPC2) in the $F606W$-band (\lameff\ = 5734\AA, \dlam\ = 2178\AA) \citep[GO-6356; P.I. M. Rowan-Robinson,][]{Farrah2001}.
The WFPC2 $F606W$-band images have the pixel scale of 0.045\arcsec\ pixel$^{-1}$ and the PSF of 0.06\arcsec\, corresponding to the physical scale of $\sim150$ pc pixel$^{-1}$ and $\sim200$ pc at the median redshift of the QDOT sample ($z=0.2$).
We stress the importance of including these 23 (U)LIRG and the necessity of using the {\it HST} $F606W$-band images due to the following reasons:
(1) the QDOT (U)LIRGs provide auxiliary sample covering the highest \LIR,
(2) the QDOT galaxies are further away compared to the GOALS galaxies, and thus {\it HST} images are required to provide comparable resolution to the high$-z$ observations,
(3) our comparison high$-z$ (U)LIRGs are observed in $F814W$-band, which corresponds to rest-frame $B$- to $U$-band at $z\sim1$, and thus we use the images with shortest possible wavelengths that are available for the local sample.
The GOALS-SDSS and QDOT samples together provide 82 (U)LIRGs (hereafter the local sample).

\begin{figure*}
 \centering
  \includegraphics[width=\textwidth]{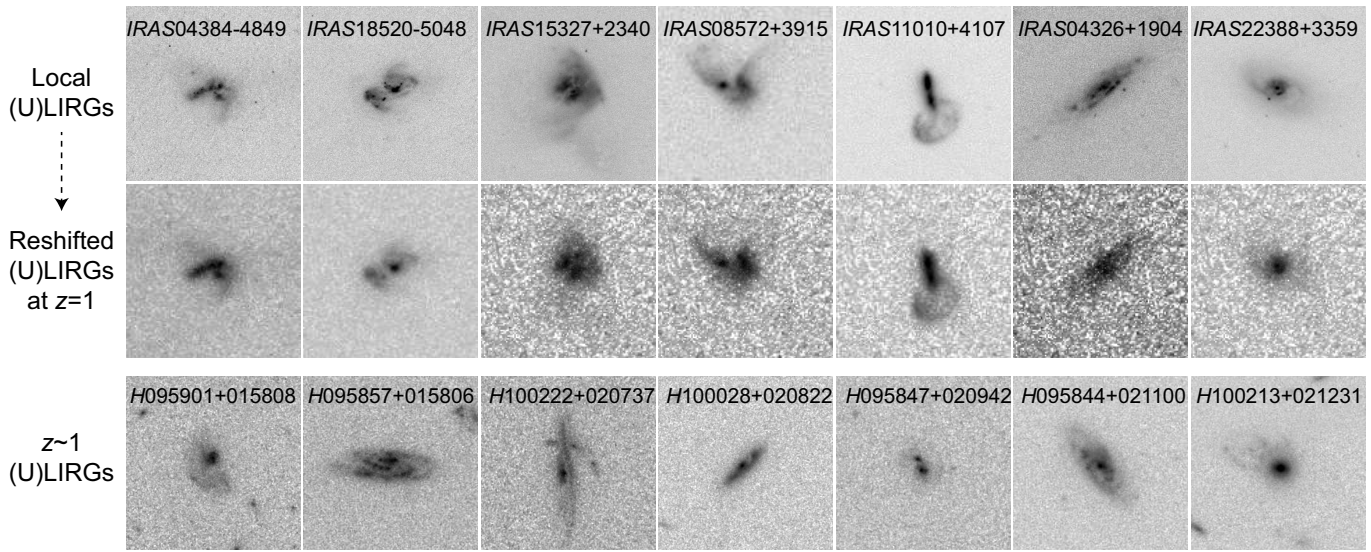} 
\caption{Examples of local, redshifted and high$-z$ (U)LIRGs.
The top two rows show the one-to-one example of local (U)LIRGs and their artificially redshifted counterparts at $z=1$ ({\it IRAS} ID of these sources is shown in each panel).
The bottom row shows examples of high$-z$ (U)LIRGs ({\it Herschel} ID of these sources is shown in each panel).
All images have the size corresponding to a physical size of 50$\times$50 kpc.
} 
\label{fig:redexample}
\end{figure*}

\section{Redshifted Local (U)LIRGs}

We have constructed a redshifted dataset of local (U)LIRGs that mimics the data quality of our high$-z$ sample based on the optical images described in Section 2.2 (hereafter the redshifted sample).
We begin by making image cutouts centered at the position of each (U)LIRG, where the angular size of each image corresponds to 50 kpc $\times$ 50 kpc in physical scale. 
Three galaxies in our local sample are widely separated pairs (with typical separation of $\sim40$ pc), and we use the cutout sizes of 80 kpc $\times$ 80 kpc.
We note that these widely separated non-interacting pairs are likely to be classified as isolated galaxies without including redshift information at high$-z$.
Then we mask out foreground stars and background galaxies in each image, and replace the masked regions by random background area in the same image (hereafter the ``cleaned images'').
The cleaned images are then artificially redshifted using a redshifting code, FERENGI \citep{Barden2008}.

Here we briefly summarize the redshifting procedure of the GOALS-SDSS sample.
The input multi-wavelength cleaned images are first deconvolved using the point spread function (PSF) of SDSS.
The redshifiting code then derives the surface brightness distribution of each galaxy according to the redshift and the plate scales of the input and output images.
The feasible redshift range of this redshifting experiment is determined by the wavelength coverage of SDSS and the desired output wavelength. 
Here we assume the redshifted (U)LIRGs are located at $z=1$, and they are observed in ACS $F814W$-band with the sensitivity the same as COSMOS ACS images (5 $\sigma$ point source limiting magnitude of 27.2). 
Using the {\sc k\_correct} routine, FERENGI determines the best-fit rest-frame SED template based on the multi-wavelength input images and then calculate the expected flux in the $F814W$-band. 
Finally, the output flux distribution is convolved with the PSF of ACS, and a noise frame is added using a blank region extracted from COSMOS {\it HST} observations. 
Since the majority of these GOALS-SDSS galaxies become too faint to study after redshifting, we apply an artificial brightening factor of two magnitudes to ensure that the redshifted dataset has similar apparent magnitude as our high$-z$ sample.

We apply a slightly different redshifting procedure to the QDOT sample.
Since galaxies in the QDOT sample are generally 1-2 magnitude brighter than the GOALS-SDSS sample, we do not apply artificial brightening for the QDOT sample.
We also disable the K-correction function in the FERENGI code since only single-filter images are used here.
Figure~\ref{fig:redexample} shows a few examples of the redshifted images.

\section{Morphology Indicators}

Non-parametric statistics (e.g. $Gini$ coefficient and \Mtwenty) and visual classifications are commonly used to classify mergers/non-mergers of (U)LIRGs \citep[e.g.][]{Lotz2004,Lotz2008a,Dasyra2008,Melbourne2008,Kartaltepe2010a}.
To compare the merger fractions in this study with those derived in the literature, we use both $Gini$-\Mtwenty\ and visual classification to determine the morphological properties in our local, redshifted, and high$-z$ sample.

\subsection{Gini coefficient and \Mtwenty}
The {\it Gini} coefficient \citep[$G$,][]{Gini1912} is defined in order to describe the dispersion of a given distribution, and it was introduced to describe the galaxy light distribution by \citet{Abraham2003}.
$G$ is defined in the range of $G=0-1$ by
\begin{equation}
G=\frac{1}{\overline{f}N(N-1)}\sum\limits_{i=1}^N (2i-N-1)f_i,
\end{equation}
where $\overline{f}$ is the average flux and $f_i$ is defined as the $i$-th lowest flux in a total of $N$ pixels.
In this definition, galaxies with single or multiple bright nuclei yield higher $G$ (typically $G>0.55$) whereas galaxies with smoother light distribution such as face-on spiral galaxies tend to have lower $G$ (typically $G<0.5$).

\Mtwenty\ is defined as the second-order moment of the brightest pixels that constitute 20\% of the total galaxy light, and then normalized by the second-order moment of the total galaxy light \citep[][hereafter L04]{Lotz2004}.
In its mathematical form,
\begin{equation}
M_{20}=log(\frac{\sum_i M_i}{M_{tot}}), \text{while} \sum_i f_i < 0.2 f_{tot}.
\end{equation}
$M_{tot}$ stands for the second-order moment of the total galaxy light:
\begin{equation}
M_{tot}=\sum\limits_{i=1}^N M_i=\sum\limits_{i=1}^N f_i[(x_i-x_c)^2+(y_i-y_c)^2],
\end{equation}
where $x_c$ and $y_c$ represents the the center of the galaxy, which is computed by minimizing $M_{tot}$.
Based on the definition, galaxies that show more concentrated light distribution such as elliptical galaxies tend to have smaller \Mtwenty\ (typical \Mtwenty\ $<-1.5$), yet galaxies that show more distributed bright features such as separated nuclei and star-forming clumps tend to have larger \Mtwenty\ (typical \Mtwenty\ $>-1.0$).

We derive $G$ and \Mtwenty\ based on a segmentation map that encloses the area for estimating the flux distribution.
Different methods for determining the segmentation map have been used in the literature \citep[e.g.][]{Conselice2003,Lotz2004,Abraham2007}.
Here we apply a ``quasi-Petrosian'' method developed by \citet[][also see Larson et al. (2014, in prep.)]{Abraham2007}, which defines the isophotal threshold as
 \begin{equation}
\eta=\frac{f_i}{<\sum \limits_{j=1}^i f_i>},
\end{equation}
where $f_i$ is the $i$-th step of the cumulative flux array, and $<\sum \limits_{j=1}^i f_i>$ is the mean surface brightness in the sorted flux array.
This method does not require a circular or elliptical aperture while calculating the isophotal flux such as the methods used in \citet{Conselice2003} and L04, and therefore it provides a better estimate for galaxies with irregular shapes.

When applying a constant isophotal threshold for all galaxies, $G$ and \Mtwenty\ are not biased by the source magnitude, cosmological dimming, and the sensitivity of observations.
An isophotal threshold of $\eta=0.2$ is commonly used for typical local datasets \citep[e.g.][]{Lotz2004,Cotini2013}, and is also used for some high$-z$ observations \citep{Abraham2007,Lotz2008a,LeeB2013}.
However, a point of caution has been raised by \citet{Law2012}, who show that using $\eta=0.2$ for their high$-z$ datasets is not satisfactory because the majority of their high$-z$ galaxies have $\eta>0.2$ at a signal-to-noise ratio (SN) of 1.5.
In these cases, the segmentation maps are either forced to include many pixels that are purely noise when enforcing a fixed threshold of $\eta=0.2$, or they are indeed only probing $\eta>0.2$ when limiting the calculation to SN $=1.5$.

Similar to the findings in \citet{Law2012}, we find that only 17\% of our high$-z$ sample has $\eta\leq0.2$ at SN $=1.5$ (20\% if we limit the sample to those with $m_I \leq24$).
We thus increase the isophotal threshold of our high$-z$ and redshifted sample to $\eta=0.3$, where 53\% of the galaxies have $\eta\leq0.3$ at SN $=1.5$ (limiting to those with $m_I \leq24$).
In the automatic calculations, we limit the enclosure of galaxy pixels to SN $=1.5$ if the observations are not sensitive to the isophotal threshold.
We also limit the calculation to the sources located within 3\arcsec\ from the image center for the high$-z$ sample and we do not derive $G$ and \Mtwenty\ for those galaxies with $m_I \geq$ 24 due to their faintness.

L04 find that local ULIRGs populate a distinctive region in $G$-\Mtwenty\ space compared to normal spiral and elliptical galaxies.
Therefore, they can define a merger/non-merger classification scheme based on a galaxy's location on the $G$-\Mtwenty\ plane.
In Figure 10 of L04, the interacting galaxies are classified based on the criteria $G > -0.113\times M_{20}+0.37$.
\citet[][hereafter L08]{Lotz2008a} further define a similar classification scheme ($G > -0.14\times M_{20}+0.33$ for interacting galaxies) for galaxies at higher redshift ($0.2<z<1.2$).
In this paper, we adopt the merger/non-merger classification criteria of L04 for our local sample, and we use the criteria in L08 for the redshifted and high$-z$ samples.
Since these criteria are derived using $\eta=0.2$, we have to convert the equation in L08 to one that is applicable for the calculation based on $\eta=0.3$. 
We apply a similar approach as \citet{Law2012}.
The majority of our local sample can achieve $\eta\leq0.2$ at SN $=1.5$, and thus we calculate $G$ and \Mtwenty\ of our local sample using both $\eta=0.2$ and $\eta=0.3$ and fit the best conversions of $G$ and \Mtwenty\ calculated using two thresholds:
\begin{equation}
G_{\eta=0.2}=0.63\times G_{\eta=0.3}+0.28, 
\end{equation}
\begin{equation}
M_{20,\eta=0.2} = 0.91\times M_{20,\eta=0.3}-0.26.
\end{equation}
Although the segmentation maps used in L04 and L08 are determined based on Petrosian radius with an elliptical aperture, which is different from the quasi-Petrosian method used in our work, the differences in the derived $G$ and \Mtwenty\ are small \citep{Law2012} and thus we do not apply another conversion.

Using the classification criteria described above, we show the fraction of interacting galaxies, non-interacting galaxies, and optically faint ($m_I>24$) galaxies in our local, redshifted, and high$-z$ sample in Figure~\ref{fig:gmbar} and \ref{fig:gmbar2}.
We note that Figure~\ref{fig:gmbar2} shows the results based on $G_{\eta=0.2}$ and $M_{20,\eta=0.2}$, and this is for a direct comparison with the results in L08 (see the discussion in Section 6.1).

\subsection{Visual Classification}

We adopt the visual classification scheme used in \citet{Hung2013}, which is a modified version of the method developed by \citet{Kartaltepe2014} \citep[also see ][]{Kartaltepe2012,Kocevski2012}.
A detailed description of this visual classification scheme and examples are presented in Section 3.2 and Figure 2 in \citet{Hung2013}.

We enlisted eight classifiers (hereafter the Team) to independently classify all galaxies.
The classifiers assign the morphology class and interaction class of each galaxy.
The morphology class includes disk, spheroid, irregular and unclassifiable.
Unless deemed ``unclassifiable'', galaxies can be given multiple morphology classes, such as an irregular disk or a spheroid with irregular features.
The interaction class includes merger, interacting pair (major or minor), non-interacting pair (major or minor) or non-interacting galaxy.
The classifiers assign only one interaction class to each galaxy unless it is noted as unclassifiable.

To simplify these classification results yet retain the information related to the interacting features and other non-interacting galaxy types, we re-categorize the morphology and interaction classes into six types: interacting system (I), minor-interacting pair (m), pure spheroid galaxy (S), non-interacting pair (P), non-interacting disk galaxy (D), and unclassifiable galaxy (U).
We then combine the classification results from the Team by majority vote.
For each galaxy, we assign it as one of the six types if equal or more than half of the classifiers agree, whereas we assign the type as ``inconclusive (C)'' if no conclusions can be reached among the Team (none of the six types exceed 50\% of votes or two types each has 50\% vote)
Figure~\ref{fig:visualexam} shows examples of six morphology types in three datasets: local, redshifted, and high$-z$ (U)LIRGs.
Figure~\ref{fig:visbar} shows the fraction of these morphology types in the three samples.

\begin{figure}
 \centering
  \includegraphics[width=0.45\textwidth]{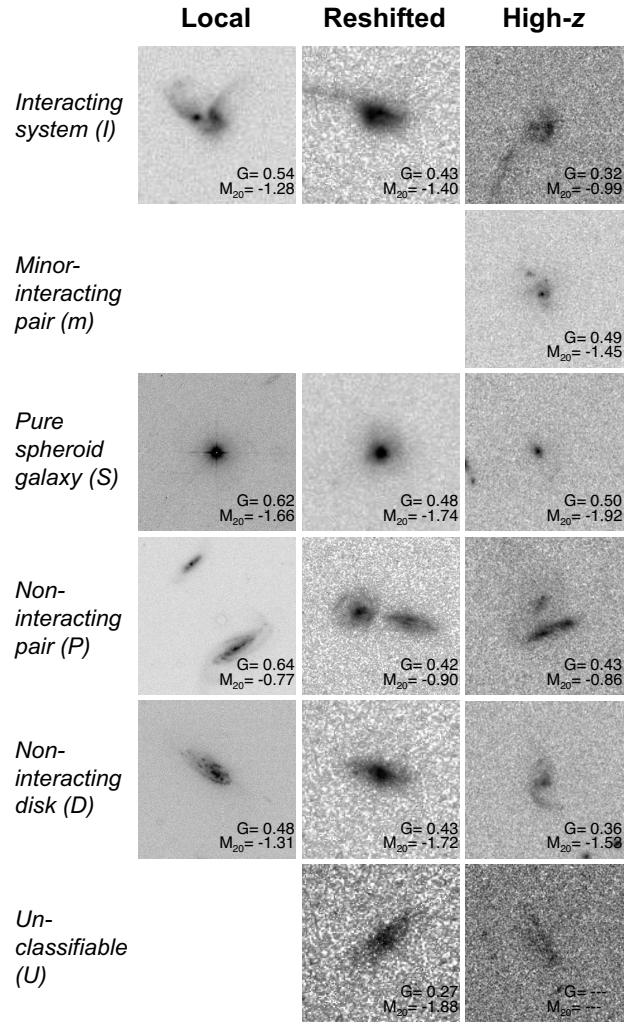} 
\caption{Examples of morphology types from our visual classifications in three datasets: local, redshifted, and high$-z$ (U)LIRGs.
The bottom-right corner of each panel lists their $G$ and \Mtwenty\ values.
All images have the size corresponding to a physical size of 50 kpc $\times$ 50 kpc.
} 
\label{fig:visualexam}
\end{figure}

\section{Results}
\subsection{Rest-frame optical morphology of local, redshifted, and high$-z$ (U)LIRGs}
In this section, we examine the distribution of morphology types for local, redshifted, and high$-z$ (U)LIRGs using both $G$-\Mtwenty\ and visual classifications.
We divide the high$-z$ sample into two redshift bins: $z=0.5-1.0$ and $z=1.0-1.5$, in which the ACS $F814W$-band images correspond to rest-frame $B$- and $U$-band in these two redshift bins.   
The distribution of morphology types is shown in three luminosity bins: Log(\LIR/\Lsun)$>12$, Log(\LIR/\Lsun)$=11.5-12$, and Log(\LIR/\Lsun)$=11-11.5$ (Figures~\ref{fig:gmbar} and ~\ref{fig:visbar}).
The error bars in these figures are determined assuming a Poisson distribution.
Typically, each luminosity and redshift bin contains at least 20-30 galaxies.

Figure~\ref{fig:gmbar} shows the classification results based on $G$-\Mtwenty.
The fractions of interacting and non-interacting systems are consistent between the local and redshifted datasets in the two higher \LIR\ bins, and differ by $\sim10\%$ in the lowest \LIR\ bin.
In fact, 77\% of the local (U)LIRGs retain their morphology classifications after redshifting.
The $G$-\Mtwenty\ classifications for the local (U)LIRGs derived with $\eta=0.2$ are highly consistent with those of the redshifted counterparts derived with $\eta=0.3$ because we define the conversions between $\eta=0.2$ and $\eta=0.3$ based on our local sample.
On average, only $6^{+14}_{-6}$\% fewer galaxies are classified as interacting systems in the redshifted sample compared to the local sample.
The overall merger fraction in the $z=0.5-1.0$ (U)LIRGs is $\sim8\%$ lower than the local sample, which is consistent with what might be expected due to redshifting (i.e. a combination effect from surface brightness dimming and degrading the spatial resolution).
The small difference in the merger fraction is consistent with an absence of strong evolution in morphological properties of (U)LIRGs out to $z\sim1$.
In the higher redshift bin ($z=1.0-1.5$), the merger fraction decreases by $\sim30\%$ in the two highest luminosity bins compared to the local sample.
However, the merger fraction at $z>1$ is highly uncertain due to the growing population of optically faint galaxies.
About 30\% of (U)LIRGs in the $z=1.0-1.5$ bin are too faint to measure $G$ and \Mtwenty\ reliably and thus their morphology types can not be determined.

The distribution of different morphology types based on visual classifications is shown in Figure~\ref{fig:visbar}.
The decrease of interacting systems in the redshifted dataset is more prominent compared to the results based on $G$-\Mtwenty\ classification.
A significant population (on average, $15^{+10}_{-8}\%$) of local (U)LIRGs lose their interacting features and they are ``misclassified'' as other morphology types when artificially redshifted to $z=1$ using the data quality of the COSMOS ACS observations.
Similar to our conclusions based on $G$-\Mtwenty\ classification, at least part of the differences of merger fraction between local and $z=0.5-1.0$ (U)LIRGs can be explained as a redshifting effect.
The remaining difference of interaction fraction is $\sim$17\%, which is consistent with the increasing population of galaxies with inconclusive morphology type.
The fraction of non-interacting disks and pairs, on the other hand, remains the same in the local and $z=0.5-1.0$ bins.
In the $z=1.0-1.5$ bin, galaxies with inconclusive morphology types and unclassifiable galaxies dominate the entire sample.

\begin{figure}
 \centering
  \includegraphics[width=0.5\textwidth]{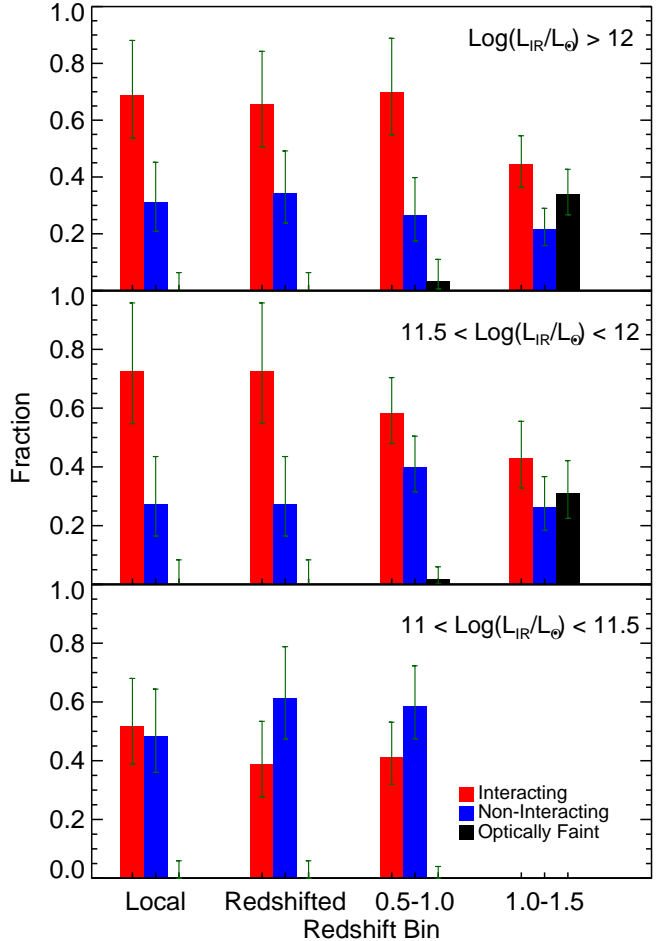} 
\caption{Distribution of morphology types based on $G-$\Mtwenty.
The adopted isophotal thresholds for the local, redshifted, and high$-z$ datasets are $\eta=0.2$, 0.3, and 0.3, respectively
Each sample is divided into three \LIR\ bins shown from top to bottom panels.
Three vertical bars indicate the fraction of interacting galaxies (red), non-interacting galaxies (blue), and the optically faint sources (black).
The error bars shown in dark green are determined assuming a Poisson distribution. 
} 
\label{fig:gmbar}
\end{figure}

\begin{figure}

  \includegraphics[width=0.5\textwidth]{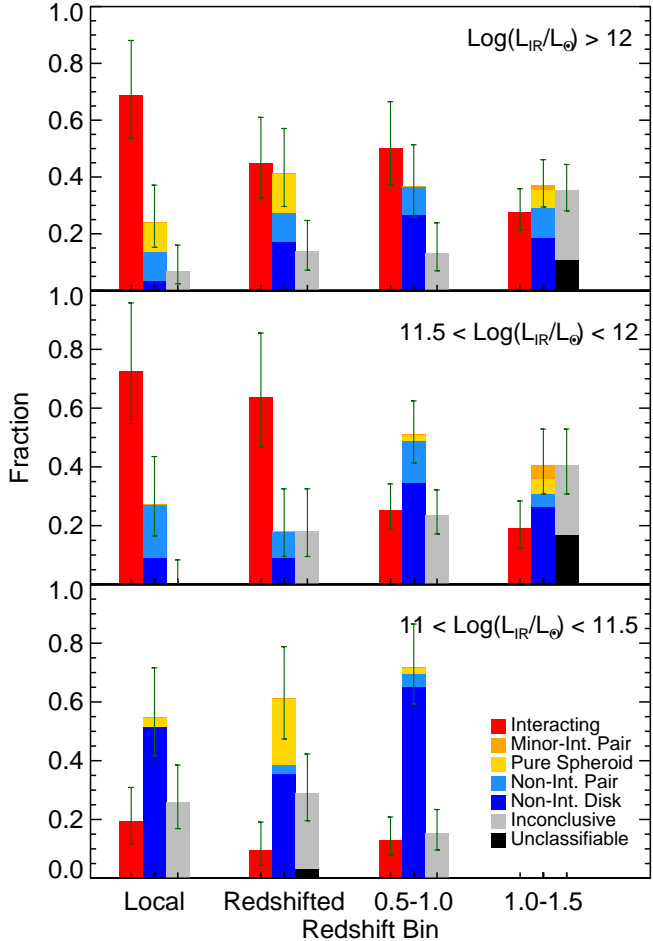} 
\caption{Distribution of morphology types based on visual classification.
Each sample is divided into three \LIR\ bins shown from top to bottom panels.
Three vertical bars indicate the fraction of one or multiple morphology types.
The left bar shows the fraction of interacting galaxies (red).
The middle bar shows the sum of minor-interacting pairs (orange), pure spheroid galaxies (yellow), non-interacting pairs (light blue), and non-interacting disks (blue).
The right bar shows the sum of inconclusive (gray) and unclassifiable (black) cases.
The error bars shown in dark green are determined assuming a Poisson distribution. 
} 
\label{fig:visbar}
\end{figure}

\subsection{Comparison between $G$-\Mtwenty\ and visual classification}
We compare the classification results based on $G$-\Mtwenty\ and visual classification in Table 1, and the distributions of $G$, \Mtwenty\ overlaid with the visual interacting systems are shown in Figure~\ref{fig:ginim20}.
The outcomes of visual classification are divided into seven possibilities (I: Interacting system, m: minor-interacting pair, S: pure spheroid, P: non-interacting pair, D: non-interacting disk, U: unclassifiable, C: inconclusive), and the outcomes of $G$-\Mtwenty\ are divided into three possibilities (interacting, non-interacting, and optically faint galaxy).
The cells with bold text indicate the cases where $G$-\Mtwenty\ and visual classifications are consistent, and we exclude the visually inconclusive cases (numbers with brackets) when calculating the level of consistency between $G$-\Mtwenty\ and visual classifications.
We note that although the visually non-interacting pairs lack of obvious interacting features, they are likely to be classified as interacting systems using $G$-\Mtwenty\ due to the presence of multiple nuclei.
Here we consider this case as the two classification results being consistent, and discuss other sources where $G$-\Mtwenty\ and visual classifications are inconsistent.

\begin{figure*}
  \centering
  \includegraphics[width=\textwidth]{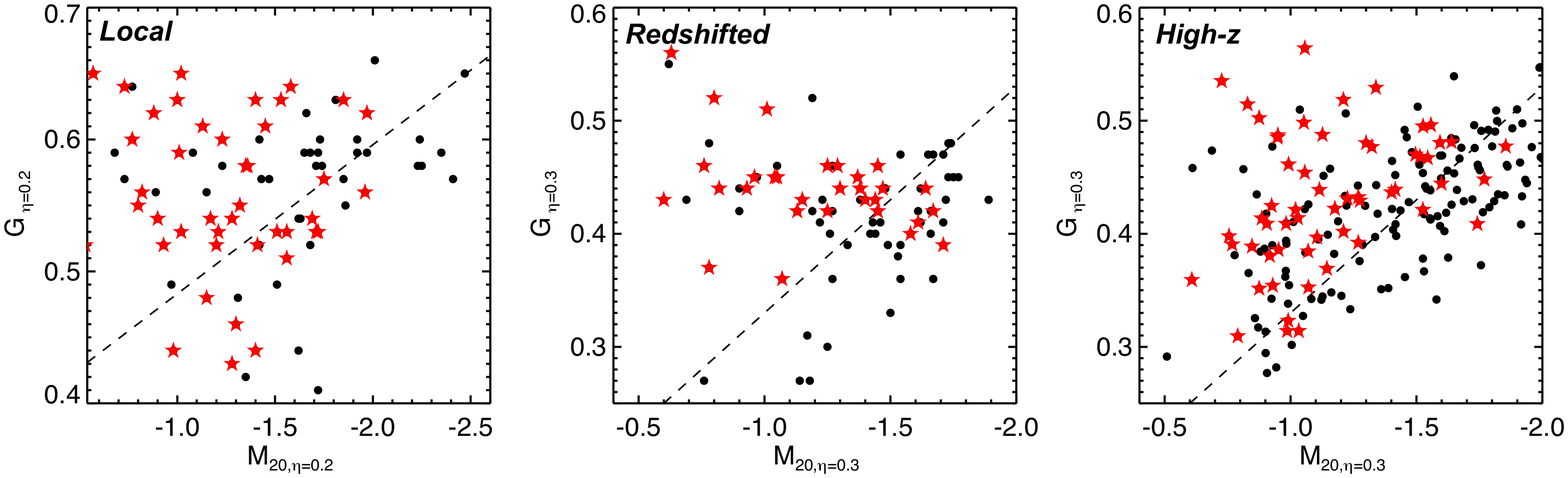} 
\caption{Distribution of $G$ and \Mtwenty\ in three datasets: local, redshifted, and high$-z$ (U)LIRGs. 
The black dots and the red stars are all galaxies with $G$-\Mtwenty\ classifications; the red stars indicate those galaxies classified as interacting systems using visual classifications.
The dotted lines are the merger/non-merger classification criteria defined in Section 4.1. 
} 
\label{fig:ginim20}
\end{figure*}

After removing the inconclusive galaxies in visual classifications, we find that the levels of consistency are 68, 74, and 61\% in the local, redshifted, and high$-z$ datasets, respectively.
Based on the local sample, we identify two major sources where the results based on $G$-\Mtwenty\ and visual classifications are inconsistent.
The galaxies with disturbed morphology that still have a relatively smooth light distribution (e.g. mergers in the interaction class) often have low $G$ value and tend to be classified as non-interacting systems using $G$-\Mtwenty.
On the other hand, edge-on disks that also have concentrated light distribution may be mistaken as interacting systems in $G$-\Mtwenty\ due to their relatively high $G$ value.
The high$-z$ dataset shows a slightly lower level of consistency compared to the local and redshifted datasets, which is due to different definitions of galaxy pairs between $G$-\Mtwenty\ and visual classifications (with and without a fixed radius).

About $12\%$ of the local sample do not have conclusive morphology type based on visual classifications, and this fraction increases by a factor of 1.6 in the redshifted and high$-z$ samples ($\sim19$ and $20\%$).
About half of these visually inconclusive galaxies are classified as interacting when using $G$-\Mtwenty\ classification, and the other half are classified as either non-interacting or optically faint.
The majority of the inconclusive cases are due to the identification of obvious interacting features, and about 10\% of the inconclusive cases in the high$-z$ sample are due to the faintness of galaxies (i.e. if galaxies are classifiable).


\begin{table}

\vspace{8pt}
{\bf Local (U)LIRGs}\\
\begin{tabular}{|l|*{7}{c|}}\hline
\backslashbox{$G$-\Mtwenty}{Visual}
&\makebox[1.5em]{I}&\makebox[1.5em]{m} & \makebox[1.5em]{S} & \makebox[1.5em]{P} & \makebox[1.5em]{D} & \makebox[1.5em]{U} & \makebox[1.5em]{C} \\\hline\hline
Interacting & \textbf{29 }&      \textbf{0}   &    3   &    \textbf{7}    &   7 &      0   &    (6)   \\\hline
Non-Interacting &13   &    0  &     \textbf{1}  &     0   &   \textbf{12} &       0   &    (4)\\\hline
Optically Faint &0 &      0  &     0  &     0  &     0 &      \textbf{0} &      (0)\\\hline
\end{tabular}
\hspace*{10pt}Total number of galaxies = 82 (72 without ``inconclusive (C)'')\\
\hspace*{140pt}Level of consistency = 68\%\\

\vspace{8pt}
{\bf Reshifted (U)LIRGs}\\
\begin{tabular}{|l|*{7}{c|}}\hline
\backslashbox{$G$-\Mtwenty}{Visual}
&\makebox[1.5em]{I}&\makebox[1.5em]{m} & \makebox[1.5em]{S} & \makebox[1.5em]{P} & \makebox[1.5em]{D} & \makebox[1.5em]{U} & \makebox[1.5em]{C} \\\hline\hline
Interacting &\textbf{24}   &    \textbf{0}    &   6  &    \textbf{ 6}  &     4 &      0  &     (7)\\\hline
Non-Interacting & 6   &    0   &    \textbf{5} &     0  &    \textbf{14}  &     1    &   (9)\\\hline
Optically Faint &0 &      0  &     0  &     0  &     0 &      \textbf{0} &      (0)\\\hline
\end{tabular}
\hspace*{10pt}Total number of galaxies = 82 (66 without ``inconclusive (C)'')\\
\hspace*{140pt}Level of consistency = 74\%\\

\vspace{8pt}
{\bf high$-z$ (U)LIRGs}\\
\begin{tabular}{|l|*{7}{c|}}\hline
\backslashbox{$G$-\Mtwenty}{Visual}
&\makebox[1.5em]{I}&\makebox[1.5em]{m} & \makebox[1.5em]{S} & \makebox[1.5em]{P} & \makebox[1.5em]{D} & \makebox[1.5em]{U} & \makebox[1.5em]{C} \\\hline\hline
Interacting & \textbf{49}  &     \textbf{3}  &     2 &     \textbf{10} &     29   &    2   &   (26)\\\hline
Non-Interacting & 8   &    0   &   \textbf{ 4 }  &    7  &    \textbf{47} &      2  &    (15)\\\hline
Optically Faint & 5 &      0  &     2  &     6  &     6  &    \textbf{13}   &   (10)\\\hline
\end{tabular}
\hspace*{5pt}Total number of galaxies = 246 (195 without ``inconclusive (C)'')\\
\hspace*{140pt}Level of consistency = 61\%\\

\textsc{  Table 1.---} Comparison tables between visual classification and the $G$-\Mtwenty\ classification results of local, redshifted, and high$-z$ datasets, including seven possible visual classification outcomes (I: interacting, m: minor-interacting, S: pure spheroid, P: non-interacting pair, D: non-interacting disk, U: unclassifiable, C: inconclusive) and three possible outcomes of $G$-M20 classification.
The tables list the number of galaxies in each cell as belonging to the certain category classified visually and using automatic indicators. 
For illustration, numbers in boldface indicate those categories where the two classification schemes are consistent, whereas normal text indicates where the two schemes are inconsistent.
The visually inconclusive cases (marked as numbers in brackets) are not included when calculating the level of consistency between the two classification schemes.

\label{tab:gmvis_com}

\end{table}

\section{Discussion}

Based on the comparison of merger fractions between local and $z\sim1$ (U)LIRGs determined using their rest-frame optical morphology, we can begin to study how the relative importance of galaxy interactions in the luminous infrared galaxies evolves with redshift.
In this section, we compare our results with previous morphological studies, and then we discuss the implications of our results.

\subsection{Comparing merger fractions to other studies in the literature}
Extensive studies based on optical morphological properties have demonstrated that merger fractions systematically increase with \LIR, and this trend is seen from the local universe out to $z\sim1$ \citep[e.g.][]{Ishida2004,Kartaltepe2010a,Hung2013}.
We observe similar increasing trends with \LIR\ in our local and high$-z$ (U)LIRGs using both $G$-\Mtwenty\ and visual classification methods (Figure~\ref{fig:gmbar} and ~\ref{fig:visbar}).
The increasing trend based on $G$-\Mtwenty\ is less prominent compared to visual classifications, which is likely due to the higher fraction of non-disturbed edge-on disks in the lowest luminosity bin (as discussed in section 5.2), and the fact that $G$-\Mtwenty\ is only sensitive to interacting galaxies at certain merger stages \citep[e.g. after the first passage and later stages,][]{Lotz2008}.

In spite of the agreement in the general trend with \LIR, some discrepancies are seen in the measured merger fractions in this study and the literature, especially for the local ULIRGs.
A slightly lower merger fraction is measured using $G$-\Mtwenty\ compared to the number in \citet{Lotz2004} (70\% v.s. 75\%), yet this difference may not be significant considering the typical uncertainties based on the sample size ($\sim15\%$).

Local ULIRGs are almost universally ($>90\%$) classified as mergers based on visual classifications \citep[e.g.][]{Sanders1996,Farrah2001,Veilleux2002,Ishida2004}, however, only $\sim70\%$ of local ULIRGs are classified as interacting systems based on our visual classification results, which is $\sim$20-30\% lower than values in the literature.
One reason for this discrepancy is the definition of the visual classification scheme.
In the classification scheme used in this paper, the detection of obvious interacting features is required for galaxies to be classified as interacting systems, and we have not incorporated available spectroscopic information to identify interacting galaxies without obvious disturbed morphology.
However, the process of ``downgrading'' the available information for the local dataset is essential to carry out a fair comparison with the high$-z$ dataset.
Another reason for the discrepancy is due to the use of optical images at shorter rest-frame wavelengths.
For example, two ULIRGs with inconclusive morphology types are identified as interacting systems when 
we reclassify the morphology types using $F814W$-band images.

The merger fractions in high$-z$ (U)LIRGs based on visual classifications are consistent with previous studies at $0.5<z<1.5$ using similar sample selection and classification scheme \citep[e.g.][]{Kartaltepe2010a,Hung2013}.
Both \citet{Kartaltepe2010a} and \citet{Hung2013} also identify a large population of galaxies with unknown morphology types at $z>1$.
In fact, the merger fraction in our $0.5<z<1.0$ LIRGs is also consistent with morphological studies based on 24 \mum-selected LIRGs at a similar redshift range \citep[$\sim$30\% at $z\sim0.7$ and $0.6<z<1.0$,][]{Bell2005,Melbourne2005}, although a lower merger fraction is also reported in some cases \citep[$\sim13\%$ at $z\sim0.8$,][]{Melbourne2008}.

The merger fraction of our $z\sim1$ LIRGs determined using $G$-\Mtwenty\ is significantly higher compared to \citet{Lotz2008a}, where they only identify $\sim15\%$ of LIRGs at $0.2<z<1.2$ as mergers.
The main reason for this discrepancy is due to the derivation of $G$ and \Mtwenty.
In Figure~\ref{fig:gmbar2}, we show that the merger fraction decreases significantly in high$-z$ (U)LIRGs based on $G_{\eta=0.2}$ and $M_{20,\eta=0.2}$ with the merger classification criteria in L08 ($G > -0.14\times M_{20}+0.33$).
However, the merger fraction also decreases significantly in the redshifted dataset, which suggests a larger correction factor due to the redshifting effect is needed when comparing local and high$-z$ datasets.

\begin{figure}
  \includegraphics[width=0.5\textwidth]{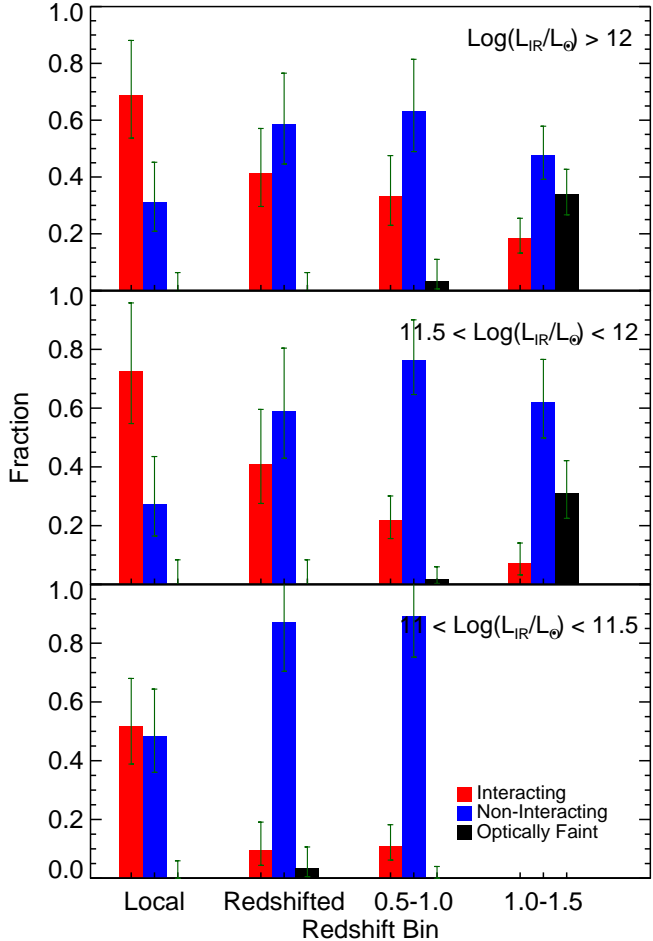} 
\caption{Distribution of morphology types based on $G-$\Mtwenty.
The adopted isophotal thresholds are $\eta=0.2$ for the local, redshifted, and high$-z$ datasets.
Each sample is divided into three \LIR\ bins shown from top to bottom panels.
Three vertical bars indicate the fraction of interacting galaxies (red), non-interacting galaxies (blue), and optically faint sources (black).
The error bars shown in dark green are determined assuming a Poisson distribution. 
} 
\label{fig:gmbar2}
\end{figure}

\subsection{Evolution of (U)LIRGs from local to $z\sim1$}

By comparing morphological studies of infrared luminous galaxies from local to higher redshifts, a general trend is seen such that the merger fraction of $z\sim1-2$ (U)LIRGs is significantly lower compared to their local counterparts \citep[by $\sim40-50\%$,][]{Ishida2004,Dasyra2008,Kartaltepe2010a}.
However, we demonstrate that the difference in merger fraction between local and $z\sim1$ (U)LIRGs may be overestimated because
(1) a lower merger fraction in local (U)LIRGs is measured when applying the same classification scheme as other studies of $z\sim1-2$ galaxies, and 
(2) the differences between local and $z\sim1$ (U)LIRGs can be partly explained as a redshifting effect.
In fact, based on our visual classification results, the merger fraction of (U)LIRGs at $z=0.5-1.0$ is only $\sim$17\% lower than local (U)LIRGs when properly accounting for (1) and (2).

A number of theoretical and empirical approaches have been proposed to explain and predict the decreasing merger fraction for $z\sim1-2$ (U)LIRGs.
For example, \citet{Hopkins2010} suggest that due to enhanced gas fractions in galaxies, the luminosity threshold between normal star formation and merger-driven starburst can increase by an order of magnitude from $z\sim0$ to $z\sim2$.
\citet{Sargent2012} show a similar increase of that luminosity threshold based on an empirical approach that assumes an increasing specific star formation rate (sSFR=SFR/\Mstar) with redshift and a double-Gaussian decomposition of the sSFR distribution at fixed stellar mass \citep{Rodighiero2011}.

The general trends of increasing merger fraction with \LIR\ and decreasing merger fraction with redshift found in this study are consistent with the prediction in \citet{Sargent2012}.
However, when taking a subset of our sample at $z=0.8-1.2$, our measured merger fraction is much higher compared to their predicted contribution of starbursts at $z=1$ (e.g. 31\% v.s. 14\% at log(\LIR/\Lsun)$\sim$11.7).
The difference may be even larger when considering the correction factor due to the redshifting effect. 
At a slightly lower redshift slice ($0.65\leq z \leq0.75$), \citet[][Figure 3]{Bell2005} also show that interacting galaxies contribute  $\sim32\%$ of the luminosity function at log(\LIR/\Lsun)$\sim$11.7.

The discrepancy between the measured merger fractions and the predicted starburst contributions at $z\sim1$ may be due to several reasons.
Firstly, this comparison assumes that the morphologically identified mergers are all starburst galaxies, which may not be valid when SFRs are only enhanced during a certain period of the interaction sequence \citep[e.g.][]{Mihos1996}.
Secondly, using asymmetry \citep[$A$,][]{Conselice2003}, \citet{Lotz2010} find that the merger observability timescale increases by a factor of two for an increase in gas fraction from 19\% to 39\%.
This implies that more mergers may be observed at $z\sim1-2$.
However, this conclusion may be highly sensitive to the merger classification scheme as \citet{Lotz2010} also point out that such trend does not exist when using $G$-\Mtwenty.

\section{Conclusions}
We have carried out a comparison of optical morphological properties between local (U)LIRGs and (U)LIRGs at $z=0.5-1.5$.
Both our local and high$-z$ samples are selected based on rest-frame 60-100 \mum\ observations, and the same classification scheme is adopted using optical images at similar rest-frame wavelengths.
In parallel, we investigate possible systematic biases when comparing two different datasets by constructing a redshifted dataset using local (U)LIRGs, in which the redshifted dataset has the same image resolution/sensitivity as the COSMOS {\it HST}-ACS observations of our high$-z$ sample.

Our determination of the distribution of merger fractions in our local, redshifted, and $z\sim1$ samples leads to the following conclusions:
\begin{enumerate}
\item Using a modified $G$-\Mtwenty\ classification scheme, no significant difference is seen in the fraction of interacting systems between local, redshifted, and (U)LIRGs at $z=0.5-1.0$ ($63^{+10}_{-9}$, $57^{+9}_{-8}$, and $55^{+7}_{-6}\%$, respectively).

\item Since we adopt a ``conservative'' visual classification scheme where the detection of obvious interacting features is required for galaxies to be classified as interacting systems, the merger fraction of local ULIRGs measured in this study is only $\sim$70\%, which is $\sim$20-30\% lower compared to the numbers reported in the literature.

\item We quantify the reduction in merger fraction that is purely due to a redshifting effect (i.e. poorer data quality at high$-z$), which is less obvious using $G$-\Mtwenty\ ($6^{+14}_{-6}\%$) but more prominent using visual classification ($15^{+10}_{-8}\%$).

\item Based on the conclusion bullet points 2 and 3, we conclude that the merger fraction of (U)LIRGs decreases by at most $\sim$20\% from $z\sim0$ to $z\sim1$.
However, $\sim$ 30\% of galaxies at $z>1$ becomes too faint for reliable classifications, and thus the merger fractions at $z>1$ suffer from large uncertainties.

\end{enumerate}

\acknowledgments
We thank the referee for her/his constructive comments that have helped us to improve the quality of this paper. C-LH wishes to acknowledge the funding support from NASA Grant NNX11AB02G, NNX14AJ61G, and the Smithsonian Astrophysical Observatory Predoctoral Fellowship. DBS and KLL gratefully acknowledge support from NASA Grant NNX11AB02G, and NL acknowledges support from SAO Grant AR3-14011X. 

Part of the data presented in this paper has been made available by the COSMOS team (http://cosmos.astro.caltech.edu). COSMOS is based on observations with the NASA/ESA Hubble Space Telescope, obtained at the Space Telescope Science Institute, which is operated by AURA Inc, under NASA contract NAS 5-26555; also based on data collected at : the Subaru Telescope, which is operated by the National Astronomical Observatory of Japan; the XMM-Newton, an ESA science mission with instruments and contributions directly funded by ESA Member States and NASA; the European Southern Observatory, Chile; Kitt Peak National Observatory, Cerro Tololo Inter-American Observatory, and the National Optical Astronomy Observatory, which are operated by the Association of Universities for Research in Astronomy, Inc. (AURA) under cooperative agreement with the National Science Foundation; the National Radio Astronomy Observatory which is a facility of the National Science Foundation operated under cooperative agreement by Associated Universities, Inc ; and the Canada-France-Hawaii Telescope operated by the National Research Council of Canada, the Centre National de la Recherche Scientifique de France and the University of Hawaii.

This paper has made use of the NASA/IPAC Extragalactic Database (NED) and IPAC Infrared Science Archive, which are operated by the Jet Propulsion Laboratory, California Institute of Technology, under contract with the National Aeronautics and Space Administration. Some of the data presented in this paper were obtained from the Mikulski Archive for Space Telescopes (MAST). STScI is operated by the Association of Universities for Research in Astronomy, Inc., under NASA contract NAS5-26555. Support for MAST for non-HST data is provided by the NASA Office of Space Science via grant NNX13AC07G and by other grants and contracts. This work has also made use of data products from the Sloan Digital Sky Survey Data Release 7. Funding for the SDSS and SDSS-II has been provided by the Alfred P. Sloan Foundation, the Participating Institutions, the National Science Foundation, the U.S. Department of Energy, the National Aeronautics and Space Administration, the Japanese Monbukagakusho, the Max Planck Society, and the Higher Education Funding Council for England. The SDSS Web Site is http://www.sdss.org/.

\bibliographystyle{apj}
\bibliography{Cosmos}

\end{document}
